\documentclass[letterpaper]{IEEEtran}
\usepackage[T1]{fontenc}
\usepackage[latin9]{inputenc}
\usepackage{color}
\usepackage{float}
\usepackage{amsmath}
\usepackage{amssymb}
\usepackage{graphicx}
\usepackage{ifpdf}
\usepackage{setspace}
\usepackage[bookmarks=true,bookmarksnumbered=true,bookmarksopen=true,bookmarksopenlevel=1,
 breaklinks=false,pdfborder={0 0 0},pdfborderstyle={},backref=false,colorlinks=false]
 {hyperref}
\hypersetup{pdftitle={Your Title},
 pdfauthor={Your Name},
 pdfpagelayout=OneColumn, pdfnewwindow=true, pdfstartview=XYZ, plainpages=false}

\makeatletter

\providecommand{\tabularnewline}{\\}
\floatstyle{ruled}
\newfloat{algorithm}{tbp}{loa}
\providecommand{\algorithmname}{Algorithm}
\floatname{algorithm}{\protect\algorithmname}

\usepackage{graphics}
\usepackage{algorithmic}
\usepackage{algorithm}
\usepackage[compatibility=false]{caption}
\usepackage{tabularray}

\usepackage{amsmath,amsfonts}
\usepackage{fancyhdr}
\ifdefined\showcaptionsetup
 \PassOptionsToPackage{caption=false}{subfig}
\fi
\usepackage{subfig}
\makeatother

\begin{document}
\title{Directivity Enhancement of Movable Antenna Arrays with Mutual Coupling}
\author{{\normalsize Wei Xu, }{\normalsize\textit{Graduate Student Member,
IEEE,}}{\normalsize{} Lipeng Zhu, }{\normalsize\textit{Senior Member, IEEE,}}{\normalsize{}
Wenyan Ma, }{\normalsize\textit{Graduate Student Member, IEEE,}}{\normalsize{}
An Liu, }{\normalsize\textit{Senior Member, IEEE, }}{\normalsize and
Rui Zhang, }{\normalsize\textit{Fellow, IEEE}}\vspace{-0.2in}
{\normalsize\thanks{This work was supported in part by the National Key Research and Development
Program of China under Grant 2025ZD1301900; in part by the Zhejiang
Provincial Key Laboratory of Information Processing, Communication
and Networking (IPCAN), Hangzhou, China. (Corresponding authors: Lipeng
Zhu; Wenyan Ma.)\par W. Xu and A. Liu are with the College of Information
Science and Electronic Engineering, Zhejiang University, Hangzhou
310027, China (e-mails: 12231077@zju.edu.cn and anliu@zju.edu.cn).\par
Lipeng Zhu is with the State Key Laboratory of CNS/ATM and the School of Interdisciplinary Science, Beijing Institute of Technology, Beijing 100081, China (E-mail: zhulp@bit.edu.cn).\par
W. Ma, and R. Zhang are with the Department of Electrical and Computer
Engineering, National University of Singapore, Singapore 117583 (e-mails:
wenyan@u.nus.edu, elezhang@nus.edu.sg).}}}
\maketitle
\begin{spacing}{0.86}
\begin{abstract}
In conventional antenna arrays, mutual coupling between antenna elements
is often regarded as detrimental. However, under specific conditions,
it can be harnessed to enhance the far-field directivity (i.e., beamforming
gain). Theoretically, the directivity of an $N$-antenna superdirective
array over the endfire direction can reach $N^{2}$, significantly
exceeding the directivity of a traditional uncoupled array which is
$N$ over all directions. This paper investigates the potential of
mutual coupling effects in movable antenna (MA) arrays for directivity
enhancement. A low-complexity algorithm called Greedy Search and Gradient
Descent (GS-GD) is proposed to optimize the antenna positions for
maximizing the array directivity over any given direction, where the
antenna positions are first selected sequentially from discrete grid
points and then continuously refined through gradient descent (GD)
optimization. Numerical results demonstrate that the optimized MA
array design by exploiting the antenna coupling achieves significant
directivity gains compared to the conventional uniform linear array
(ULA) without antenna coupling over all directions. Additionally,
the proposed GS-GD algorithm is shown to approach the global optimum
closely in most directions.
\end{abstract}


\section{Introduction}
\thispagestyle{empty}In recent years, multiple-input multiple-output
(MIMO) technology has achieved remarkable success in wireless communication
systems. By increasing the number of antennas, MIMO systems can enhance
both beamforming and multiplexing gains, thereby significantly improving
the achievable transmission rate. However, the physical
size constraints of modern devices and/or infrastructures necessitate
placing multiple antennas in close proximity, which inevitably leads
to mutual coupling between antenna elements. Such mutual coupling
distorts the signals and is usually regarded to be harmful to the system
performance. Numerous studies aim to eliminate or mitigate
mutual coupling based on physical isolation or signal processing
techniques \cite{gGain2025,voMutual2018}. Meanwhile, several
recent works indicate that coupled arrays can improve the far-field
directivity (i.e., beamforming/array gain in a line-of-sight scenario)
by appropriately adjusting the effective radiation pattern
\cite{marzettaSuperDirective2019a,pizzoSuperdirectivity2024}.
For example, for an $N$-antenna uniform linear array (ULA) with the
inter-antenna spacing approaching zero (i.e., superdirective array
\cite{hanSuperdirective2024}), the directivity over the endfire direction
(i.e., the direction aligned with the linear array) can reach $N^{2}$
\cite{marzettaSuperDirective2019a,pizzoSuperdirectivity2024}. In
comparison, the directivity for an uncoupled ULA with half-wavelength
spacing (i.e., ULAH) is always $N$ over all directions. This phenomenon
of coupled arrays is known as superdirectivity. Although superdirective
arrays can achieve higher beamforming gains over the endfire direction,
they suffer from substantially lower gain over other directions compared
to uncoupled arrays. In other words, the directivity of an antenna
array depends on the array geometry (i.e., antennas' positions) and
its maximum value varies with the wave direction. Thus, the maximum
directivity over all wave directions cannot be simultaneously achieved
by any array with fixed-position antennas (FPAs).
\textcolor{black}{Recently, movable antenna (MA) array has been proposed
to enhance wireless communication performance by exploiting the degrees
of freedom (DoFs) in antenna position optimization \cite{zhuModeling2024,zhuMovable2024a,maCompressed2023b}.
Joint antenna position and rotation design is also considered in six-dimensional
MA (6DMA) systems \cite{shao6D2025,shao6D2025a}. However, most existing
studies assume that no mutual coupling exists between MA elements
by constraining the inter-antenna spacing being greater than half-wavelength
\cite{zhuMovableAntenna2023}. }While prior works have demonstrated
through simulations that mutual coupling can enhance the performance
of dipole MA or reconfigurable intelligent surface (RIS) systems \cite{mursiaT3DRIS2025,zhuMutual2026},
to the best of our knowledge, a theoretical framework quantifying
the fundamental beamforming gain of general MA systems remains absent.
\textcolor{black}{To this end, this paper takes an isotropic MA array
as a case study to establish a systematic framework for harnessing
mutual coupling to enhance MA array directivity. In this paper, we
first provide the expression of the directivity to illustrate the
increased beamforming gain obtained by the coupled MA array. Then,
for any given direction, we formulate an optimization problem to maximize
the directivity by designing the antenna positions within the movable
region. To solve the optimization problem, we propose a low-complexity
Greedy Search and Gradient Descent (GS-GD) al}gorithm, where the antenna
positions are first selected sequentially from discrete grid points
and then continuously refined through gradient descent (GD) optimization.
Extensive simulation results demonstrate the directivity gain of the
coupled MA array and the efficiency of the proposed GS-GD algorithm.

\textit{Notations}: $\boldsymbol{0}_{N}$ and $\boldsymbol{I}_{N}$ refer to an all-zero column
vector of dimension $N$ and an $N\times N$ identity matrix, respectively.
$\overline{\left(\cdot\right)}$, $\left(\cdot\right)^{\mathrm{T}}$,
$\left(\cdot\right)^{\mathrm{H}}$, $\left(\cdot\right)^{\mathrm{-1}}$
and $\Re\left(\cdot\right)$ denote conjugate, transpose, conjugate
transpose, inverse and real part, respectively. $\mathbb{R}$ and
$\mathbb{C}$ refer to the sets of real numbers and complex numbers,
respectively. $|a|$ and $||a||_{2}$ denote the amplitude of scalar
$a$ and the 2-norm of vector $\boldsymbol{a}$, respectively. $a_{n}$
and $A_{mn}$ refer to the $n$-th entry of vector $\boldsymbol{a}$
and the $\left(m,n\right)$-th entry of matrix $\boldsymbol{A}$,
respectively. $\mathcal{O}\left(\cdot\right)$ denotes the order of
complexity. $o\left(\cdot\right)$ denotes higher-order infinitesimals.
\end{spacing}
\begin{spacing}{0.92}
\section{System Model and Performance Characterization}

\subsection{Directivity Model of the Coupled MA Array}

As illustrated in Fig. \ref{fig:systemmodel}, we consider a linear
MA array with $N$ isotropic antenna elements located along the $x$-axis.
The antenna position vector (APV) is denoted by $\boldsymbol{x}=\left[x_{1},x_{2},\cdots,x_{N}\right]^{\mathrm{T}}\in\mathbb{R}^{N}$.
We assume that the antennas can move on a line segment of length $d_{\max}$,
and the minimum spacing between antennas is $d_{\min}$, i.e.,
\begin{equation}
d_{\min}\leq|x_{m}-x_{n}|\leq d_{\max},\;1\leq m\neq n\leq N.\label{eq:dmax_constraint}
\end{equation}
{\small{}
\begin{figure}[t]
\vspace{-8pt}
\begin{centering}
{\small\includegraphics[width=0.8\columnwidth,height=0.25\columnwidth]{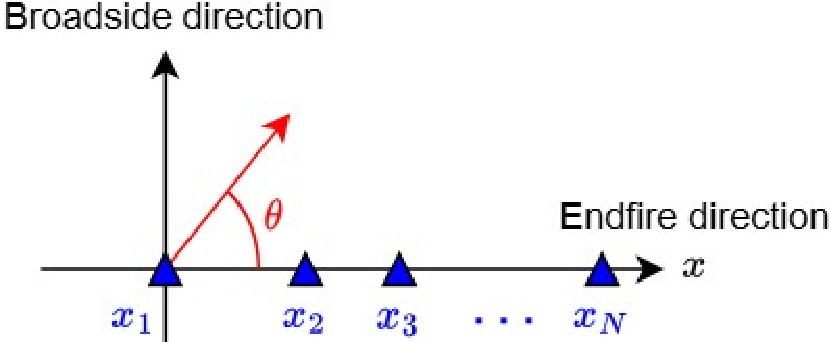}}{\small\par}
\par\end{centering}
{\small\caption{\label{fig:systemmodel}Illustration of the considered linear MA array.}
}{\small\par}
\vspace{-8pt}
\end{figure}
} The complex far-field pattern of the MA array over the direction
$\theta\in\left[0{^\circ},180{^\circ}\right]$ is expressed as a function
of $u\triangleq\cos\theta\in\left[-1,1\right]$ for notation simplicity, where $u$ is the direction
cosine and is proportional to the discrete spatial frequency \cite{pizzoSuperdirectivity2024}:
\[
y\left(u,\boldsymbol{x}\right)=\boldsymbol{w}^{\mathrm{H}}\boldsymbol{a}\left(u,\boldsymbol{x}\right),
\]
where $\boldsymbol{a}\left(u,\boldsymbol{x}\right)\in\mathbb{C}^{N\times1}$
is the steering vector (SV):
\[
\boldsymbol{a}\left(u,\boldsymbol{x}\right)=\left[e^{-j\frac{2\pi}{\lambda}x_{1}u},e^{-j\frac{2\pi}{\lambda}x_{2}u},\cdots,e^{-j\frac{2\pi}{\lambda}x_{N}u}\right]^{\mathrm{T}},
\]
with $\lambda$ denoting the wavelength. The vector $\boldsymbol{w}\in\mathbb{C}^{N\times1}$
represents the coupled excitation coefficients, where the $n$-th
entry $w_{n}$ is proportional to the current excitation on the $n$-th
antenna. Then, the total radiated power of the MA array is given by
the integral in the angular domain \cite{hanSuperdirective2024}:
\begin{equation*}
P_{rad}=\frac{1}{2}\int_{0}^{\pi}|\boldsymbol{a}^{\mathrm{H}}\left(\cos\theta,\boldsymbol{x}\right)\boldsymbol{w}|^{2}\sin\left(\theta\right)d\theta=\boldsymbol{w}^{\mathrm{H}}\boldsymbol{R}\left(\boldsymbol{x}\right)\boldsymbol{w},
\end{equation*}
where we define\textcolor{black}{
\[
\boldsymbol{R}\left(\boldsymbol{x}\right)\triangleq\frac{1}{2}\int_{0}^{\pi}\boldsymbol{a}\left(u,\boldsymbol{x}\right)\boldsymbol{a}^{\mathrm{H}}\left(u,\boldsymbol{x}\right)du,
\]
which is also called impedance coupling matrix caused by power radiation
interaction between antennas \cite{hanSuperdirective2024,pizzoSuperdirectivity2024},
and the $\left(m,n\right)$-th entry of $\boldsymbol{R}\left(\boldsymbol{x}\right)$
that represents the impedance coupling coefficient between the $m$-th
antenna and the $n$-th antennas is $R_{mn}\left(\boldsymbol{x}\right)=\frac{1}{2}\int_{-1}^{1}a_{m}\left(u,x_{m}\right)\overline{a_{n}\left(u,x_{n}\right)}du=\mathrm{sinc}\left(2\frac{x_{n}-x_{m}}{\lambda}\right)$,
with $\mathrm{sinc}\left(x\right)=\frac{\sin\left(\pi x\right)}{\pi x}$.}

\textcolor{black}{Accordingly, the directivity of the MA array over
the direction $\theta$, expressed in terms of $u$, takes the form
of a Rayleigh quotient $\mathring{G}_{u}\left(\boldsymbol{x},\boldsymbol{w}\right)=\frac{|\boldsymbol{a}^{\mathrm{H}}\left(u,\boldsymbol{x}\right)\boldsymbol{w}|^{2}}{\boldsymbol{w}^{\mathrm{H}}\boldsymbol{R}\left(\boldsymbol{x}\right)\boldsymbol{w}}$.
For a given $\boldsymbol{x}$, the optimal normalized $\boldsymbol{w}$
for maximizing the directivity is given by $\boldsymbol{w}^{*}=\frac{\boldsymbol{R}^{-1}\left(\boldsymbol{x}\right)\boldsymbol{a}\left(u,\boldsymbol{x}\right)}{||\boldsymbol{R}^{-1}\left(\boldsymbol{x}\right)\boldsymbol{a}\left(u,\boldsymbol{x}\right)||_{2}}$,
yielding the maximum directivity over $u$ as a function of the APV:
\begin{equation}
G_{u}\left(\boldsymbol{x}\right)=\boldsymbol{a}^{\mathrm{H}}\left(u,\boldsymbol{x}\right)\boldsymbol{R}^{-1}\left(\boldsymbol{x}\right)\boldsymbol{a}\left(u,\boldsymbol{x}\right).\label{eq:wstar-1-1}
\end{equation}
In most existing works for MIMO system design, impedance coupling
is neglected by constraining the inter-antenna spacing being greater
than half-wavelength, since the sinc function becomes negligible for
arguments larger than 1. Under this assumption, the radiated power,
optimal beamforming vector, and maximum directivity simplify to $P_{rad}=\boldsymbol{w}^{\mathrm{H}}\boldsymbol{w}$,
$\boldsymbol{w}^{*}=\boldsymbol{a}(u,\boldsymbol{x})/\sqrt{N}$, and
$G_{u}=N$, respectively.}

\subsection{\textcolor{black}{Directivity Gain Analysis\label{subsec:Directivity-Gain-Analysis}}}

\textcolor{black}{Direct analysis of the relationship between APV
and directivity in (\ref{eq:wstar-1-1}) is challenging for a general
array, primarily due to the presence of the inverse coupling matrix
$\boldsymbol{R}^{-1}\left(\boldsymbol{x}\right)$. To facilitate analysis,
we interpret the coupling matrix $\boldsymbol{R}\left(\boldsymbol{x}\right)$
as the Gram matrix of the pattern functions $\left\{ a_{n}\left(u,x_{n}\right)\right\} _{n=1}^{N}$
with the inner product defined as
\[
\left\langle a_{m}\left(u,x_{m}\right),a_{n}\left(u,x_{n}\right)\right\rangle \triangleq\frac{1}{2}\int_{-1}^{1}a_{m}\left(u,x_{m}\right)\overline{a_{n}\left(u,x_{n}\right)}du.
\]
}where $a_{n}\left(u,x_{n}\right)=e^{-j\frac{2\pi}{\lambda}x_{n}u}$
is the $n$-th entry of the SV $\boldsymbol{a}\left(u,\boldsymbol{x}\right)$.
To orthogonalize these pattern functions, we apply Cholesky decomposition
denoted by $\boldsymbol{R}\left(\boldsymbol{x}\right)=\boldsymbol{L}\left(\boldsymbol{x}\right)\boldsymbol{L}^{\mathrm{H}}\left(\boldsymbol{x}\right)$.
The inverse of the resulting lower-triangular matrix, $\boldsymbol{L}^{-1}\left(\boldsymbol{x}\right)$,
actually serves as the transformation matrix that explicitly constructs
a set of orthonormal pattern functions (or effective SV) $\check{\boldsymbol{a}}\left(u,\boldsymbol{x}\right)=\left(\boldsymbol{L}\left(\boldsymbol{x}\right)\right)^{-1}\boldsymbol{a}\left(u,\boldsymbol{x}\right)$
via Gram-Schmidt process. This process proceeds orthogonormalization
recursively, with the $n$-th, $n=1,2,\cdots,N$, pattern function
$a_{n}\left(u,x_{n}\right)$ orthonormalized as:\begin{subequations}\label{eq:aorth}
\begin{align}
\tilde{a}_{n}\left(u,x_{n}\right)&=a_{n}\left(u,x_{n}\right)\nonumber\\
&-\sum_{m=1}^{n-1}\left\langle \check{a}_{m}\left(u,x_{m}\right),a_{n}\left(u,x_{n}\right)\right\rangle \check{a}_{m}\left(u,x_{m}\right)\label{eq:atilde},
\end{align}
\vspace{-12pt}
\begin{align}
\label{eq:acheck}\check{a}_{n}\left(u,x_{n}\right)&=\frac{\tilde{a}_{n}\left(u,x_{n}\right)}{\sqrt{\left\langle \tilde{a}_{n}\left(u,x_{n}\right),\tilde{a}_{n}\left(u,x_{n}\right)\right\rangle }}.
\end{align}
\end{subequations}

By defining the effective beamforming vector as $\check{\boldsymbol{w}}=\boldsymbol{L}^{\mathrm{H}}\left(\boldsymbol{x}\right)\boldsymbol{w}$,
the directivity simplifies to a canonical Rayleigh quotient form $\mathring{G}_{u}\left(\boldsymbol{x},\boldsymbol{w}\right)=\frac{|\check{\boldsymbol{a}}^{\mathrm{H}}\left(u,\boldsymbol{x}\right)\check{\boldsymbol{w}}|^{2}}{\check{\boldsymbol{w}}^{\mathrm{H}}\check{\boldsymbol{w}}}$.
The optimal effective beamforming vector is thus $\check{\boldsymbol{w}}^{*}=\frac{\check{\boldsymbol{a}}\left(u,\boldsymbol{x}\right)}{||\check{\boldsymbol{a}}\left(u,\boldsymbol{x}\right)||_{2}},$
and the corresponding maximum directivity is given by:
\begin{align}
G_{u}\left(\boldsymbol{x}\right) & =||\check{\boldsymbol{a}}\left(u,\boldsymbol{x}\right)||_{2}^{2}=\sum_{n=1}^{N}|\check{a}_{n}\left(u,x_{n}\right)|^{2},\label{eq:Gtheta}
\end{align}
which are consistent with results in (\ref{eq:wstar-1-1}). This transformation
incorporates the coupling matrix $\boldsymbol{R}\left(\boldsymbol{x}\right)$
into the effective SV $\check{\boldsymbol{a}}\left(u,\boldsymbol{x}\right)$,
yielding a mathematical form identical to that of an uncoupled array.
The key difference is that the elements in the effective SV, i.e.,
$\left\{ \check{a}_{n}\left(u,x_{n}\right)\right\} _{n=1}^{N}$, are
generally not unimodular, owing to the transformation by $\left(\boldsymbol{L}\left(\boldsymbol{x}\right)\right)^{-1}$.

This formulation reveals a fundamental insight. For any given FPA
array, the average directivity over the entire angular space is $\frac{1}{2}\int_{-1}^{1}\sum_{n=1}^{N}|\check{a}_{n}\left(u,x_{n}\right)|^{2}du$,
which is always equal to $N$ because the effective pattern functions
are orthonormal, as defined in (\ref{eq:aorth}). In particular, for
arrays with elements located at integer multiples of half-wavelength
(including ULAH), we have $\boldsymbol{R}\left(\boldsymbol{x}\right)=\boldsymbol{I}_{N}$,
$\check{\boldsymbol{a}}\left(u,\boldsymbol{x}\right)=\boldsymbol{a}\left(u,\boldsymbol{x}\right)$,
and thus the directivity remains a constant of $\sum_{n=1}^{N}|a_{n}\left(u,x_{n}\right)|^{2}=N$
over all directions. In contrast, for MA arrays, the directivity over
any direction can exceed $N$ by leveraging the coupling effect via
APV optimization, which is demonstrated by the following three examples.

\subsubsection{Example I}

We first consider a simple two-antenna case (i.e., $N=2$). We set
$x_{1}=0$ without loss of generality. Based on (\ref{eq:aorth}),
the effective pattern functions are given by
\[
\check{a}_{1}\left(u,0\right)=1,\check{a}_{2}\left(u,x_{2}\right)=\frac{a_{2}\left(u,x_{2}\right)-\mathrm{sinc}\left(2\frac{x_{2}}{\lambda}\right)}{\sqrt{1-\mathrm{sinc}^{2}\left(2\frac{x_{2}}{\lambda}\right)}}.
\]
Then the directivity over any given $u$ can be expressed as
\begin{align*}
G_{u}\left(\left[0,x_{2}\right]^{\mathrm{T}}\right) & =2\frac{1-\cos\left(2\pi\frac{x_{2}u}{\lambda}\right)\mathrm{sinc}\left(2\frac{x_{2}}{\lambda}\right)}{1-\mathrm{sinc}^{2}\left(2\frac{x_{2}}{\lambda}\right)}.
\end{align*}
Over the broadside direction (i.e., $u=0$), this simplifies to $G_{u}=\frac{2}{1+\mathrm{sinc}\left(2\frac{x_{2}}{\lambda}\right)}$.
The optimal position $x_{2}\approx0.72\lambda$ yields a maximum directivity
of $G_{u}^{*}\approx2.55$. Over the endfire direction (i.e., $u=1$),
$x_{2}\rightarrow0$ yields superdirectivity $G_{u}^{*}\rightarrow4$.
Both cases achieve a directivity larger than the number of antennas
($N=2$). However, over other directions, the optimal solution for
$x_{2}$ becomes analytically complicated. The relationship between
directivity of the two-antenna MA array and $x_{2}$ is depicted in
Fig. \ref{fig:N2}.{\small{}
\begin{figure}[t]
\begin{centering}
{\small\includegraphics[clip,scale=0.37]{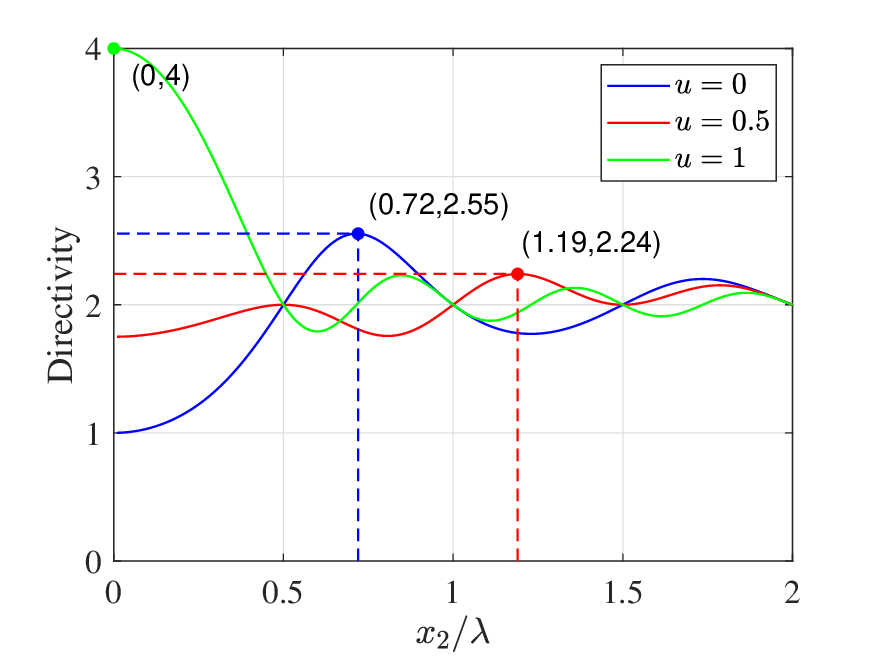}}{\small\par}
\par\end{centering}
{\small\caption{\label{fig:N2}Directivity versus $\frac{x_{2}}{\lambda}$ for $N=2$
and $u=0,0.5,1$.}
}{\small\par}
\end{figure}
}{\small\par}

\subsubsection{Example II\label{subsec:Examples-II}}

For an array with inter-antenna spacing greater than half-wavelength,\textcolor{black}{{}
}i.e., $|x_{m}-x_{n}|>\frac{\lambda}{2},\forall m\neq n$, the coupling
effects are weak. For \textcolor{black}{convenience, we define $k_{mn}=\frac{2\pi\left(x_{m}-x_{n}\right)}{\lambda},\forall m\neq n$,
and the magnitude of coupling coefficient $\left|\mathrm{sinc}\left(\frac{k_{mn}}{\pi}\right)\right|<0.22$
since $\frac{k_{mn}}{\pi}>1$. By neglecting higher-order terms $o\left(\mathrm{sinc}\left(\frac{k_{mn}}{\pi}\right)\right),\forall m\neq n$,
the orthogonalization in }(\ref{eq:aorth}) can be simplified as follows:
\begin{align*}
\check{a}_{1}\left(u,0\right) & \equiv1,\check{a}_{2}\left(u,x_{2}\right)\approx-\mathrm{sinc}\left(\frac{k_{12}}{\pi}\right)+a_{2}\left(u,x_{2}\right),\cdots,
\end{align*}
\textcolor{black}{
\[
\check{a}_{N}\left(u,x_{N}\right)\approx-\sum_{n=1}^{N-1}\mathrm{sinc}\left(\frac{k_{nN}}{\pi}\right)e^{-jk_{nN}u}+a_{N}\left(u,x_{N}\right).
\]
Under this first-order approximation, the squared magnitude of each
orthogonalized pattern simplifies to
\begin{equation}
|\check{a}_{n}\left(u,x_{n}\right)|^{2}\approx1-2\sum_{m=1}^{n-1}\mathrm{sinc}\left(\frac{k_{mn}}{\pi}\right)\cos\left(k_{mn}u\right).\label{eq:aicheck}
\end{equation}
}To validate this approximation, we consider an ULA with element spacing
$d>\frac{\lambda}{2}$, i.e., $x_{n}=\left(n-1\right)d$. The behavior
of the approximated $|\check{a}_{n}\left(u,\left(n-1\right)d\right)|^{2}$
in (\ref{eq:aicheck}) as a function of $\frac{d}{\lambda}$ is illustrated
in Fig. \ref{fig:N_app}.{\small{}
\begin{figure}[t]
\vspace{-10pt}
\begin{centering}
{\small\subfloat[$n=2$.]{\begin{centering}
{\small\includegraphics[clip,scale=0.27]{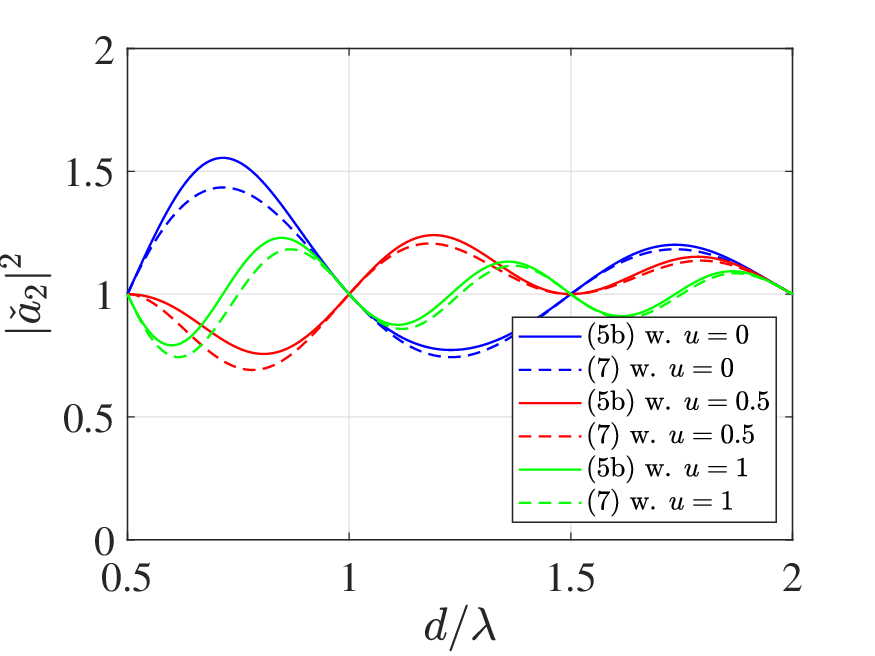}}{\small\par}
\par\end{centering}
{\small}{\small\par}}}{\small\hspace{0.01\textwidth}}{\small\subfloat[$n=3$.]{\begin{centering}
{\small\includegraphics[clip,scale=0.27]{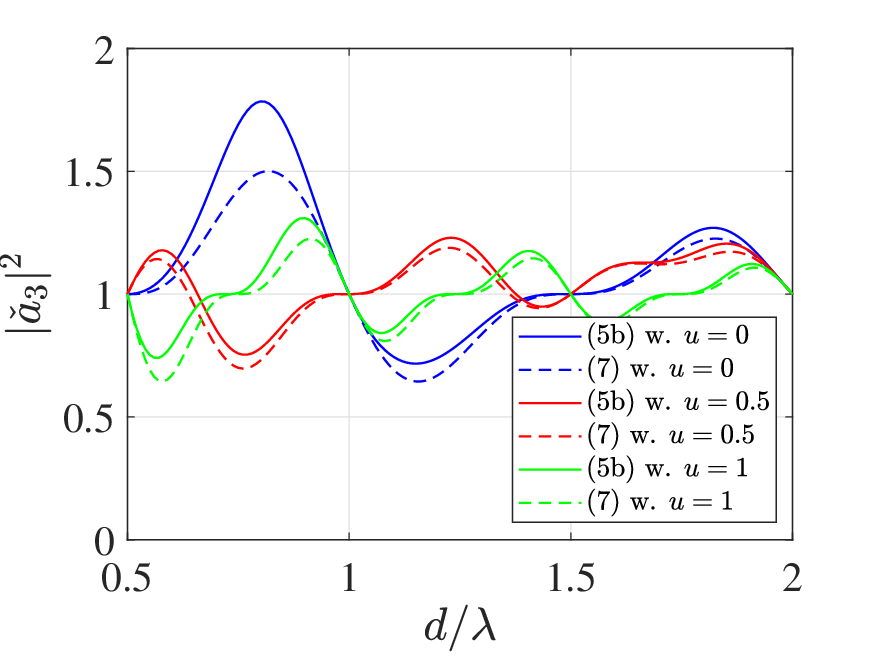}}{\small\par}
\par\end{centering}
{\small}{\small\par}}}{\small\par}
\par\end{centering}
{\small\caption{\label{fig:N_app}For an ULA with $d>\frac{\lambda}{2}$, approximation
of $|\check{a}_{n}\left(u,x_{n}\right)|^{2}$ via (\ref{eq:aicheck})
against squared magnitude of (\ref{eq:acheck}) versus $d$ for $n=2,3$
and $u=0,0.5,1$.}
}{\small\par}
\vspace{-10pt}
\end{figure}
}{\small\par}

\textcolor{black}{Consequently, the overall directivity $G_{u}\left(\boldsymbol{x}\right)$
is given by}
\begin{align*}
 & \sum_{n=1}^{N}|\check{a}_{n}\left(u,x_{n}\right)|^{2}\approx N-2\sum_{1\leq m<n\leq N}\mathrm{sinc}\left(\frac{k_{mn}}{\pi}\right)\cos\left(k_{mn}u\right).
\end{align*}
A special case arises for the broadside direction ($u=0$). In this
case, maximizing the directivity is equivalent to minimizing the sum
$\sum_{1\leq m<n\leq N}\mathrm{sinc}\left(\frac{k_{mn}}{\pi}\right)$.
If only mutual coupling between adjacent elements is considered, the
directivity can be further simplified as $G_{u}\left(\boldsymbol{x}\right)\approx N-2\sum_{n=2}^{N}\mathrm{sinc}\left(2\frac{x_{n}-x_{n-1}}{\lambda}\right)$.
Under this simplified model, the optimal APV reduces to an ULA with
$d\approx0.72\lambda$, yielding a directivity of
\begin{equation}
G_{u}\left(\boldsymbol{x}\right)\approx1.44N-0.44.\label{eq:di_app}
\end{equation}

The directivities over the broadside direction for ULAs with $d\approx0.72\lambda$
under different values of $N$ are shown in Table \ref{tab:dapp}.{\small{}
\begin{table}[tbh]
\begin{centering}
\begin{tabular}{|c|c|c|c|c|}
\hline 
$N$ & 2 & 3 & 4 & 5\tabularnewline
\hline 
Directivity in (\ref{eq:Gtheta}) & 2.55 & 4.13 & 5.49 & 6.88\tabularnewline
\hline 
Approximation in (\ref{eq:di_app}) & 2.44 & 3.88 & 5.32 & 6.76\tabularnewline
\hline 
\end{tabular}
\par\end{centering}
\centering{}{\small\caption{\label{tab:dapp}Approximated directivity in (\ref{eq:di_app}) against
actual directivity in (\ref{eq:Gtheta}) under different values of
$N$.}
}{\small\par}
\end{table}
}{\small\par}

For arrays with elements $N\geq3$, the directivity over the broadside
direction can be further enhanced by fine-tuning antenna positions
to exploit mutual coupling effects from non-adjacent elements.

\subsubsection{Example III}

For a superdirective array, i.e., $x_{n}=\left(n-1\right)d$ with
$d\rightarrow0$, we define $k=\frac{2\pi d}{\lambda}\rightarrow0$
for convenience, and the pattern functions orthogonalized via (\ref{eq:atilde})
are given by
\[
\tilde{a}_{1}\left(u,0\right)\equiv1,\tilde{a}_{2}\left(u,d\right)=-jku+o\left(k\right),
\]
\[
\tilde{a}_{3}\left(u,2d\right)=\frac{2}{3}k^{2}+2k^{2}u^{2}+o\left(k^{2}\right),\cdots,
\]
where higher-order terms $o\left(k^{n-1}\right)$ in $\tilde{a}_{n}\left(u,\left(n-1\right)d\right)$
can be neglected. After normalization, the effective pattern functions
correspond to $\check{a}_{n}\left(u,\left(n-1\right)d\right)=\sqrt{2n-1}P_{n-1}\left(u\right),n=1,2,\cdots,N$,
where $P_{n}\left(\cdot\right)$ is the $n$-th order Legendre polynomial
\cite{levinNearField2021} with $P_{0}\left(u\right)=1$,
$P_{1}\left(u\right)=u$, $P_{2}\left(u\right)=\frac{3u^{2}-1}{2}$,
etc. The directivity thus becomes $\sum_{n=1}^{N}\left(2n-1\right)|P_{n-1}\left(u\right)|^{2}$,
which attains a maximum of $N^{2}$ over the endfire direction ($u=\pm1$).

For a general $N$-antenna MA array, a closed-form solution for the
optimal APV is intractable. In the following, we aim to optimize the
APV numerically. The optimization problem can be formulated as follows:\begin{subequations}\label{eq:problem}
\begin{align}
\label{eq:G1}\underset{x_{2},x_{3},\cdots,x_{N}}{\max}&\quad G_{u}\left(\boldsymbol{x}\right)\\
\label{eq:asumpt}\mathrm{s.t.}&\quad(\ref{eq:dmax_constraint}).
\end{align}
\end{subequations}

\section{Proposed algorithm}

Since $x_{1}=0$, the rest $\left(N-1\right)$ antennas are constrained
within the movable region $\mathcal{P}=\left[-d_{\max},-d_{\min}\right]\cup\left[d_{\min},d_{\max}\right]$,
while simultaneously satisfying the constraints specified in (\ref{eq:asumpt}).
We propose a computationally efficient two-stage algorithm for optimizing
$x_{2},\cdots,x_{N}$. In the first stage, the $\left(N-1\right)$
antenna positions are selected from grid points sampled over $\mathcal{P}$
based on greedy search (GS). In the second stage, the antenna positions
are refined continuously based on GD.

\subsection{Grid Points Selection Based on GS\label{subsec:GS}}

We sample $M$ grid points $\tilde{\mathcal{X}}\triangleq\left\{ \tilde{x}_{1},\tilde{x}_{2},\cdots,\tilde{x}_{M}\right\} $
with uniform spacing $d_{g}$ ($d_{g}<d_{\min}$) over $\mathcal{P}$.
We define an index vector $\mathcal{\boldsymbol{\mathcal{M}}}\triangleq\left[\mathcal{\mathcal{M}}_{2},\mathcal{\mathcal{M}}_{3},\cdots,\mathcal{\mathcal{M}}_{N}\right]^{\mathrm{T}}$,
where each $\mathcal{M}_{n}\in\left\{ 1,2,\cdots,M\right\} $ corresponds
to the grid point index for the $n$-th antenna position.

The GS procedure optimizes these indexes sequentially. In the $n$-th
step ($n=1,2,\cdots,N-1$), the positions of the first $n$ antennas
are fixed. The algorithm then selects the optimal grid point for $x_{n+1}$
by exhaustively evaluating all candidate points in $\tilde{\mathcal{X}}$
that satisfy the position constraints in (\ref{eq:asumpt}), with
the goal of maximizing the directivity of the $(n+1)$-element array.
The GS sub-problem in the $n$-th step is thus given by\begin{subequations}\label{eq:GthetaM}
\begin{align}
\underset{\mathcal{\mathcal{M}}_{n+1}}{\max}&\quad G_{u}\left(\left[0,x_{2}^{GS},\cdots,x_{n}^{GS},\tilde{x}_{\mathcal{\mathcal{M}}_{n+1}}\right]^{\mathrm{T}}\right)\\
\label{eq:asumptGS}\mathrm{s.t.}&\quad d_{\min}\leq |\tilde{x}_{\mathcal{\mathcal{M}}_{n+1}}-x_{m}^{GS}|\leq d_{\max},2\leq m\leq n,
\end{align}
\end{subequations}which can be efficiently solved by the one-dimension (1D) exhaustive
search (ES), with the optimal solution denoted as $\mathcal{\mathcal{M}}_{n+1}^{*}$.
The selected position is fixed as $x_{n+1}^{GS}=\tilde{x}_{\mathcal{\mathcal{M}}_{n+1}^{*}}$
for subsequent steps. 

\subsection{Antenna Positions Refinement Based on GD\label{subsec:GD}}

Since the optimal positions generally do not lie exactly on the discrete
grid points in $\tilde{\mathcal{X}}$, we employ GD for continuous
refinement. Specifically, we denote the maximum iteration number and
the APV obtained in the $t$-th iteration as $T$ and $\boldsymbol{x}\left(t\right)$,
respectively. The APV is initialized as the output of the GS process
denoted as $\boldsymbol{x}^{GS}=\left[0,x_{2}^{GS},\cdots,x_{N}^{GS}\right]^{\mathrm{T}}$,
i.e., $\boldsymbol{x}\left(0\right)=\boldsymbol{x}^{GS}$. Then, the
$t$-th iteration is detailed as follows.

First, by defining $\dot{\boldsymbol{a}}\left(t\right)=\boldsymbol{R}^{-1}\left(\boldsymbol{x}\left(t\right)\right)\boldsymbol{a}\left(u,\boldsymbol{x}\left(t\right)\right)$,
the partial derivatives of the objective function $G_{u}\left(\boldsymbol{x}\right)$
with respect to (w.r.t.) $x_{n},n=2,\cdots,N$ are given by
\begin{align}
\frac{\partial G_{u}\left(\boldsymbol{x}\right)}{\partial x_{n}} & =2\Re\left(\dot{\boldsymbol{a}}^{\mathrm{H}}\left(t\right)\frac{\partial\boldsymbol{a}\left(u,\boldsymbol{x}\right)}{\partial x_{n}}\right)-\dot{\boldsymbol{a}}^{\mathrm{H}}\left(t\right)\frac{\partial\boldsymbol{R}\left(\boldsymbol{x}\right)}{\partial x_{n}}\dot{\boldsymbol{a}}\left(t\right),\label{eq:gradients}
\end{align}
where elements in $\frac{\partial\boldsymbol{a}\left(u,\boldsymbol{x}\right)}{\partial x_{n}}\in\mathbb{\mathbb{R}}^{N}$
are all zeros except the $n$-th element $\left(\frac{\partial\boldsymbol{a}\left(u,\boldsymbol{x}\right)}{\partial x_{n}}\right)_{n}=-j\frac{2\pi}{\lambda}ue^{-j\frac{2\pi}{\lambda}x_{n}\left(t\right)u}$,
and elements in $\frac{\partial\boldsymbol{R}\left(\boldsymbol{x}\right)}{\partial x_{n}}\in\mathbb{\mathbb{R}}^{N\times N}$
are all zeros except elements in the $n$-th row and elements in the
$n$-th column $\left(\frac{\partial\boldsymbol{R}\left(\boldsymbol{x}\right)}{\partial x_{n}}\right)_{mn}=\begin{cases}
0, & m=n,\\
\frac{\cos\left(\pi\left(x_{n}\left(t\right)-x_{m}\left(t\right)\right)\right)-\mathrm{sinc}\left(x_{n}\left(t\right)-x_{m}\left(t\right)\right)}{x_{n}\left(t\right)-x_{m}\left(t\right)}, & m\neq n.
\end{cases}$

Next, the APV is updated using an adaptive step size $\alpha$, initialized
as $\alpha_{0}$. A candidate update for $\boldsymbol{x}\left(t\right)$
is computed as
\begin{equation}
\boldsymbol{x}^{test}=\boldsymbol{x}\left(t\right)+\alpha\left[0,\frac{\partial G_{u}}{\partial x_{2}},\cdots,\frac{\partial G_{u}}{\partial x_{N}}\right]^{\mathrm{T}}.\label{eq:xtest}
\end{equation}
This candidate is accepted only if it satisfies two conditions: On
one hand, it improves the objective function, i.e., $G_{u}\left(\boldsymbol{x}^{test}\right)>G_{u}\left(\boldsymbol{x}\left(t\right)\right)$;
on the other hand, it satisfies the position constraints in (\ref{eq:asumpt}).
Otherwise, $\alpha$ is halved in a backtracking line search until
the conditions are satisfied or $\alpha$ is below the threshold $\epsilon$.

\subsection{Algorithm Summary}

The proposed algorithm is summarized in Algorithm \ref{GSGD}. Lines
1-4 correspond to the GS stage, while lines 5-16 correspond to the
GD stage. The convergence of GD is guaranteed because the objective
function is upper-bounded, and non-decreasing in iterations. The computational
complexity is analyzed as follows. The directivity evaluation in (\ref{eq:Gtheta})
of $M$ candidate grid points for $\mathcal{\mathcal{M}}_{n+1}$ in
line 3 can be reduced to computing $\left\{ \check{a}_{n+1}\left(u,x\right)\right\} _{x\in\tilde{\mathcal{X}}}$
with complexity $\mathcal{O}(Mn^{2})$, since $\left\{ \check{a}_{m}\left(u,x_{m}^{GS}\right)\right\} _{m=1}^{n}$
are recursively computed in former steps. The complexity of calculating
the gradients in line 8 is $\mathcal{O}\left(N^{2}\right)$. The complexity
of calculating the objective function in line 12 is $\mathcal{O}\left(N^{3}\right)$,
and thus the maximum complexity of the entire while loop is $\mathcal{O}\left(\log_{2}\left(\frac{\alpha_{0}}{\epsilon}\right)N^{3}\right)$.
Hence, the overall computational complexity of the proposed Algorithm
\ref{GSGD} is $\mathcal{O}\left(MN^{3}+T\left(N^{2}+\log_{2}\left(\frac{\alpha_{0}}{\epsilon}\right)N^{3}\right)\right)$.
{\small{}
\begin{algorithm}[t]
{\small\caption{\label{GSGD}GS-GD for Antenna Position Optimization}
}{\small\par}

{\small\textbf{Input:}}{\small{} $N,\theta,\lambda,d_{\max},d_{\min},d_{g},\alpha_{0},\epsilon$.}{\small\par}

\small
\begin{algorithmic}[1]

\STATE \%\% The first stage: GS to select the antenna positions
from grid points.

\FOR{$n=1\cdots N-1$}

\STATE Optimize $\mathcal{M}_{n+1}$ by solving (\ref{eq:GthetaM}).

\ENDFOR

\STATE \%\% The second stage: GD to continuously refine the
positions via backtracking line search.

\STATE Initialize $\boldsymbol{x}\left(0\right)=\boldsymbol{x}^{GS}$.

\FOR{$t=0\cdots T-1$}

\STATE Compute the gradients $\frac{\partial G_{u}\left(\boldsymbol{x}\right)}{\partial x_{n}},n=2,\cdots,N$
as (\ref{eq:gradients}).

\STATE Initialize $\alpha=\alpha_{0},r=-1,\boldsymbol{x}^{test}=\boldsymbol{x}\left(t\right)$.

\WHILE{$r<0$ or $\boldsymbol{x}^{test}$ does not satisfy (\ref{eq:asumpt})}

\STATE Update $\boldsymbol{x}^{test}$ as (\ref{eq:xtest}).

\STATE Compute $r=G_{u}\left(\boldsymbol{x}^{test}\right)-G_{u}\left(\boldsymbol{x}\left(t\right)\right)$.

\STATE $\alpha\leftarrow0.5\alpha$.

\IF{$\alpha<\epsilon$}

\STATE Return $\boldsymbol{x}^{*}=\boldsymbol{x}\left(t\right)$.

\ENDIF

\ENDWHILE

\STATE Update $\boldsymbol{x}\left(t+1\right)=\boldsymbol{x}^{test}$.

\ENDFOR

\STATE $\boldsymbol{x}^{*}=\boldsymbol{x}\left(T\right)$.

\end{algorithmic}

{\small\textbf{Output:}}{\small{} $\boldsymbol{x}^{*}$.}{\small\par}
\end{algorithm}
}{\small\par}

\section{Numerical results}

This section presents numerical re\textcolor{black}{sults to validate
the directivity enhancement of the MA array by taking into the antenna
coupling effect and the performance of the proposed GS-GD algorithm.
For clarity, the directivity is presented as a function of the angle
$\theta$ rather than $u$ in this section. Furthermore, the results
are confined to $\theta\in\left[0{^\circ},90{^\circ}\right]$, as
the directivity is symmetric w.r.t. $90{^\circ}$. The main simulation parameters
are summarized in Table~\ref{tab:params} for ease of reference.}

\begin{table}[t]
\centering
\begin{tabular}{|c|c|}
\hline
Parameter & Value \\
\hline
Number of antennas $N$ & 5 \\
\hline
Wavelength $\lambda$ & $0.3$ m \\
\hline
Minimum spacing $d_{\min}$ & $\lambda/10 = 0.03$ m \\
\hline
Aperture size $d_{\max}$ & $\frac{N-1}{2}\lambda$, $(N-1)\lambda$, $2(N-1)\lambda$ \\
\hline
Grid spacing $d_g$ & $\lambda/20 = 0.015$ m \\
\hline
GS-GD iterations $T$ & 5 \\
\hline
Initial step size $\alpha_0$ & 1 \\
\hline
Tolerance $\epsilon$ & $10^{-3}$ \\
\hline
\end{tabular}
\caption{\label{tab:params}Main simulation parameters.}
\end{table}

\textcolor{black}{First, we compare the directivity of the optimized
coupled MA array with the traditional ULAH. In simulations, we employ
an ES over all discretized grid points to approximate the globally
optimal solution for problem (\ref{eq:problem}). The results are
shown in Fig. \ref{fig:directivity0}.}{\small{}
\begin{figure}[t]
\begin{centering}
{\small\includegraphics[clip,scale=0.45]{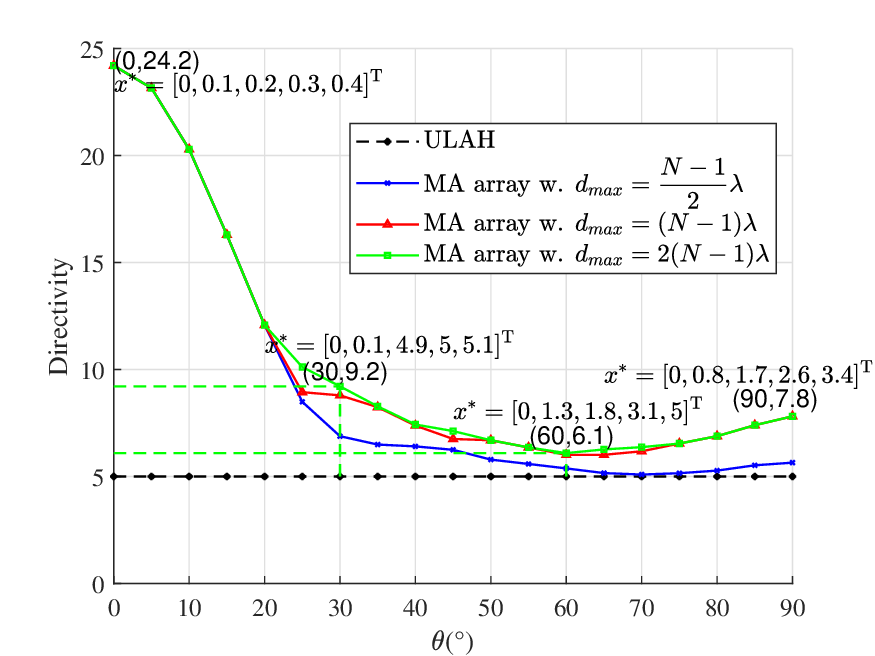}}{\small\par}
\par\end{centering}
{\small\caption{\label{fig:directivity0}Directivity of the coupled MA array compared
to the ULAH in different values of $\theta$ and $d_{\max}$.}
}{\small\par}
\vspace{-6pt}
\end{figure}
} 

The results demonstrate that, over all directions, the optimized MA
array achieves a higher directivity compared to the ULAH. Specifically,
when $\theta\rightarrow0{^\circ}$, the superdirective array is optimal,
and the directivity approaches $N^{2}$. As $\theta$ increases, the
optimal antenna positions exhibit irregular patterns, and there is
at least 20\% improvement in the directivity for the coupled MA array
compared to the traditional ULAH without antenna coupling when the
movable region is sufficiently large. When $\theta\rightarrow90{^\circ}$,
the optimal array converges to a nearly uniform array with spacing
around $0.8\lambda$, and there is around 50\% directivity gain, which
are consistent with the analysis in Section \ref{subsec:Examples-II}.
Additionally, over the vast majority of directions, the maximum directivity
for $d_{\max}=\left(N-1\right)\lambda$ is the same as that for $d_{\max}=2\left(N-1\right)\lambda$,
both of which substantially exceed the directivity for $d_{\max}=\frac{1}{2}\left(N-1\right)\lambda$,
which implies that $d_{\max}=\left(N-1\right)\lambda$ is sufficient
to achieve most of the directivity gains in practice.

Next, we evaluate the performance of the proposed GS-GD algorithm
against the following baseline methods:
\begin{itemize}
\item ES: This method selects antenna positions from grid points, and compute
the directivity of all feasible APV cases to find the maximum directivity.
\item \textcolor{black}{GS: This method only performs the first stage as
mentioned in Section \ref{subsec:GS}.}
\item \textcolor{black}{GD: This method initializes the} APV as a ULAH,
and then performs \textcolor{black}{the second stage} as mentioned
in Section \ref{subsec:GD}, where we set $T=30$ in simulations to
ensure convergence.
\item ULAH: For ULAH without antenna coupling, the directivity is $N$ over
all directions.
\end{itemize}
{\small{}
\begin{figure}[t]
\begin{centering}
{\small\includegraphics[clip,scale=0.45]{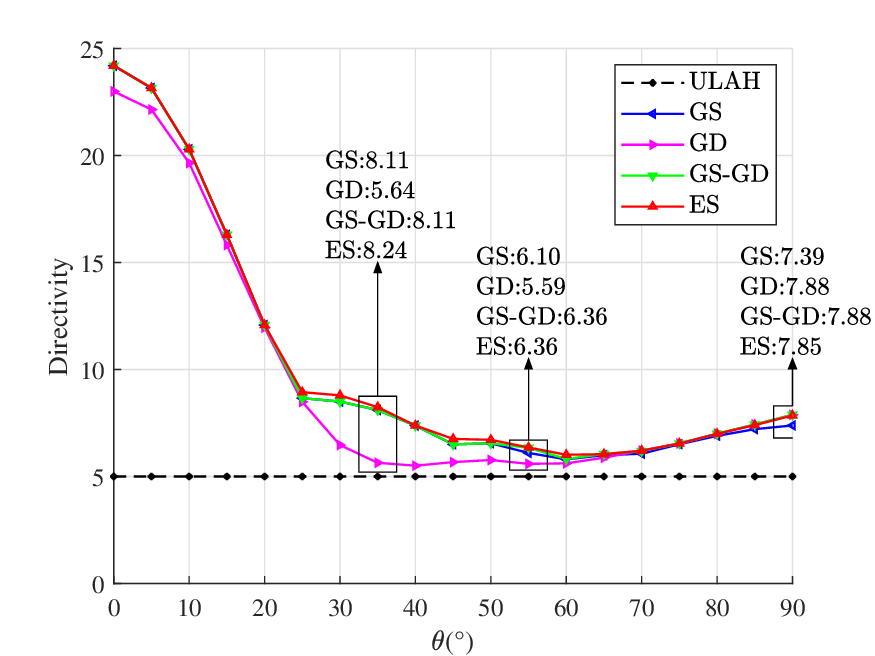}}{\small\par}
\par\end{centering}
{\small\caption{\label{fig:directivity0-1}Comparison of the proposed GS-GD algorithm
with baselines in different values of $\theta$.}
}{\small\par}
\vspace{-6pt}
\end{figure}
}{\small\par}

The performance of the algorithms is evaluated in Fig. \ref{fig:directivity0-1}.
As shown in the figure, the performance of the proposed GS-GD approaches
ES over most directions. Moreover, GS-GD outperforms the two baselines
(GS and GD) over all directions. This superiority arises from its
ability to mitigate the inherent limitations of both methods: GS is
constrained by its sequential optimization process, which myopically
adjusts individual antenna positions, thus lacking the capability
to co-optimize the entire array, while GD is prone to local optima
when initialized from a ULAH. By leveraging both approaches, GS-GD
enables a more robust search, consistently converging to a solution
closer to the global optimum.

\section{Conclusion}

In this paper, we investigated the directivity of MA arrays with antenna
coupling. In contrast to traditional uncoupled arrays, coupled MA
arrays can enhance the directivity (i.e., beamforming gain) by concentrating
the radiated power more effectively toward specific directions. We
aimed to optimize the positions of antennas for maximizing the directivity
of the MA array over any given direction. A low-complexity GS-GD algorithm
was proposed, where the antenna positions are first selected sequentially
from discrete grid points through GS and then continuously refined
through GD optimization. Numerical results demonstrated the directivity
gain of the optimized coupled MA array and the efficiency of the proposed
GS-GD algorithm.

\bibliographystyle{IEEEtran}
\bibliography{ref/MA_SU_0917,ref/DNN_CE,ref/turboVBI,ref/SU_addition_1005,ref/antenna_coupling_cancellation,ref/MA_0127}

\end{spacing}

\end{document}